\documentclass[sigconf,nonacm]{acmart}

\setcitestyle{numbers,sort&compress}
\usepackage{tabularx}
\usepackage{booktabs}
\usepackage[table]{xcolor}
\usepackage[table]{xcolor}
\usepackage{threeparttable}
\usepackage{subcaption}
\usepackage{tikz}
\usepackage{multirow}
\usepackage{float}
\usepackage{enumitem}
\usepackage{stfloats}
\usetikzlibrary{positioning}
\usepackage{xurl}
\usepackage{tabularx}
\usepackage{booktabs}
\usepackage{hyperref}
\usepackage{xcolor}
\begin{document}

\title{Student Perspectives on the Benefits and Risks of AI in Education}

\author{Griffin Pitts}
\orcid{0009-0004-3111-6118}
\authornote{Corresponding author.}
\affiliation{%
  \institution{University of Florida}
  \city{Gainesville}
  \state{Florida}
  \country{USA}
}
\email{wgpitts@ncsu.edu}
  
\author{Viktoria Marcus}
\orcid{0000-0003-1997-8108}
\affiliation{%
  \institution{University of Florida}
  \city{Gainesville}
  \state{Florida}
  \country{USA}
}
\email{viktoria.marcus@ufl.edu}
  
\author{Sanaz Motamedi}
\orcid{0009-0005-0938-9029}
\affiliation{%
  \institution{Pennsylvania State University}
  \city{University Park}
  \state{Pennsylvania}
  \country{USA}}
\email{sjm7946@psu.edu}

\renewcommand{\shortauthors}{Pitts et al.}

\begin{abstract}
The use of chatbots equipped with artificial intelligence (AI) in educational settings has increased in recent years, showing potential to support teaching and learning. However, their adoption has raised concerns about academic integrity, students' ability to problem-solve independently, and potential underlying biases. To better understand students' perspectives on these benefits and risks, a survey was conducted at a large public university in the United States, collecting 262 responses from undergraduate students. Through thematic analysis, students' perceived benefits and risks of AI chatbots in education were identified and categorized into themes. The results identified perceived benefits related to feedback and study support, instructional capabilities, and access to information. Students also expressed concerns about academic integrity, information accuracy, loss of critical thinking skills, and overreliance on AI, alongside ethical issues involving data privacy, system bias, environmental impact, and the preservation of human interaction.

While student perceptions align with previously discussed benefits of AI in education, they also highlight concerns about distinguishing between human- and AI-generated work and the broader ethical implications of these technologies. To address students' concerns regarding academic integrity and information reliability, institutions can establish clear policies regarding AI use and develop curricula around AI literacy. With these in place, practitioners can effectively develop and implement educational systems that leverage AI's potential in areas such as immediate feedback and personalized learning support. This approach can enhance the quality of students' educational experiences while preserving the integrity of the learning process with AI.
\end{abstract}


\maketitle

\colorbox{gray!10}{%
  \parbox{\dimexpr\linewidth-2\fboxsep\relax}{%
    \small
    \textbf{Published Version:}\par\smallskip
    This work was published in the proceedings of the
    \textit{2025 ASEE Annual Conference \& Exposition}. You may cite the published version as:\par\bigskip

    Griffin Pitts, Viktoria Marcus, and Sanaz Motamedi.
    Student Perspectives on the Benefits and Risks of AI in Education. In \textit{2025 ASEE Annual Conference \& Exposition}, 2025.

    \par\smallskip
    \textbf{DOI:}
    \href{https://doi.org/10.18260/1-2--57693}
    {10.18260/1-2--57693}
  }%
}

\section{Introduction and Background}

Artificial intelligence (AI) chatbots have emerged as a growing resource in educational settings. Advances in large language models (LLMs) have enhanced AI chatbots' ability to understand and respond to academic queries, driving their increased adoption in educational settings and sparking greater research interest. Open online models such as OpenAI's ChatGPT, Google's Gemini, and Microsoft's Copilot have gained widespread student adoption due to their free access and ease of use. This expansion has occurred amid varying acceptance \cite{a,b,c} and trust \cite{25} in digital learning technologies across student populations through the COVID-19 pandemic and into the present day. In prior studies, approximately one-third (35.4\%) of students reported regular usage of ChatGPT, while 47\% expressed concern about AI's impact in education \cite{31}. Additionally, 60\% reported that their instructors or schools had not yet provided guidelines for ethical or responsible AI tool use \cite{31}. 

As students have begun using available online AI assistants, researchers have concurrently developed specialized educational AI tools designed specifically for learning environments. These include `Jill Watson,' a chatbot designed to support student learning as a virtual teaching assistant implemented in classrooms at Georgia Tech \cite{28}, `Anne G. Neering,' a chatbot designed to motivate and engage students in engineering classrooms \cite{15}, and `GPTeach,' a chat-based tool that acts as a student to help train novice teaching assistants \cite{27}. There have been additional use cases for AI chatbots, and their underlying LLMs, highlighted in prior research specific to engineering education: assisting students in producing engineering spreadsheets in an advanced structural steel design course \cite{12}, generating feedback to students' written code in an introductory engineering courses \cite{11, pitts3}, and generating recommendations toward problem-solving in capstone courses \cite{3}. While the mentioned studies highlight how AI can be beneficial to students, this technology also presents challenges that need to be addressed for successful implementation. The following sections explore reported benefits and risks of AI in education.

\subsection{Benefits of AI Chatbots in Education}

Benefits of AI chatbots in educational settings extend to both students and educators. These systems can support learning by providing detailed explanations of concepts through intelligent tutoring support \cite{1,2,6,7,21,16}. AI systems can offer immediate feedback on students' work, allowing for more rapid improvement and iterative learning \cite{2,3,6,7,11,16,18,21}. These systems have demonstrated their potential to support personalized learning, adapting to individual student needs \cite{1,2,4,6,10,11,14,16,3}. Personalization naturally fosters higher levels of student engagement, found to be another benefit of these systems \cite{16,15,11,9,6,4}. Studies have also highlighted AI's potential to stimulate creativity through idea generation \cite{2,18,19,21}, bridge language barriers via content translation for multilingual learners \cite{21,3}, and democratize education by providing continuous access to educational support and resources regardless of time or geographical constraints \cite{3,6,14,16,18}.

Beyond student support, AI chatbots demonstrate potential for enhancing teaching efficiency through the automation of routine tasks and administrative responsibilities \cite{1,6,8,9,10,14,19,21}. This may enable educators to focus less on menial tasks, and more on student-instructor interactions. For example, AI chatbots' capability to generate text can enable human-in-the-loop systems \cite{19}, where instructors provide prompts to generate initial feedback that they can then customize for their students. This approach allows teachers to provide meaningful feedback while conserving their time and energy \cite{19}. Benefits to instructors additionally extend to support with educational content creation \cite{8,19}, learning analytics to supervise large classrooms \cite{18}, and automated grading and assessment \cite{1,6,7,3,19}.

\subsection{Risks of AI Chatbots in Education}

Even with potential benefits, implementing AI chatbots in educational settings requires careful consideration of several risks. For students, academic integrity has been raised as a primary concern, with studies highlighting risks intentional and unintentional plagiarism \cite{2,17,18,21,23}. These tools may inadvertently undermine students' critical thinking development and academic agency \cite{10,18}, potentially fostering an unhealthy overreliance on automated assistance \cite{2,23}. The quality of interaction itself presents additional challenges, as emotionally-inconsiderate exchanges can diminish student engagement \cite{10,14}, while the reduction of peer-to-peer and student-instructor interactions threatens to eliminate meaningful learning relationships \cite{14,21,3}. Furthermore, current AI systems' demonstrate a limited understanding of nuanced individual learning needs and educational contexts \cite{3}, potentially resulting in standardized experiences that contradict the promise of personalization.

Educators face their own set of implementation and oversight challenges when integrating AI systems into teaching practices. Varying levels of AI literacy \cite{1,10,18}, among both students and instructors, creates barriers to effective human-AI collaboration. Meanwhile, instructors shoulder the burden of verifying information accuracy \cite{10,19,51,29,25} and addressing potential biases in AI-generated content \cite{3,19,2}, complicated by the inherent lack of transparency in AI decision-making processes \cite{6,10,49,40,25}. Additional practical challenges add to these concerns, including managing of data privacy and security \cite{1,2,6,10,19,21,3,51, 40,29,25}, maintaining supervision of student-AI interactions \cite{1,6,14}, securing adequate technical infrastructure required for AI use \cite{1,6,14}, and ensuring the sustainability of these practices \cite{2}. 

\subsection{Research Questions}

Despite the range of benefits and risks identified in prior work, there remains limited understanding of how students themselves view and experience these technologies in their learning. To address this gap, this study specifically analyzes student perspectives on the benefits and risks of AI chatbot usage in educational settings. Therefore, our research questions are:
\begin{itemize}
\item {\textbf{RQ1: What do students perceive as the benefits of using AI chatbots in educational settings?}}
\item {\textbf{RQ2: What do students perceive as the risks of using AI chatbots in educational settings?}}
\end{itemize}

\section{Methodology}
\subsection{Procedure}

To answer these research questions, an online survey was conducted using Qualtrics software during the Fall 2024 academic semester, yielding 388 responses. The 388 responses were filtered to include only complete submissions from undergraduate students who addressed both open-ended questions, which resulted in a final sample of 262 participants. The survey structure began with demographic questions and items related to students' prior AI usage to characterize the participant sample. Participants were then provided with a clear definition of AI chatbots through an introductory video to ensure a common understanding of the technology being discussed. Finally, two open-ended questions were asked: ``If any, what do you see as the key benefits or opportunities for using AI chatbots in educational settings?” and ``If any, what do you see as the main risks or concerns around using AI chatbots in educational settings?” Following these sections, the survey contained additional questions that addressed separate research objectives not covered in the current study.

\subsection{Participants}

Participants for this study were recruited from undergraduate classrooms at the University of Florida through researcher contacts. Of the 262 participants, 129 (49.2\%) were male, 122 (46.6\%) were female, with the remaining 11 (4.2\%) choosing not to disclose or self-describing gender. 62.9\% identified as White, 27.4\% as Hispanic or Latino/a, 15.1\% as Asian, 8.5\% as Black, 4.2\% as Middle Eastern or North African, 2.3\% preferred not to disclose, and 0.8\% as Pacific Islander. Regarding prior experience with AI chatbots, 90.4\% of participants indicated they had previously used AI chatbots (with 66.0\% expressing strong familiarity and 24.4\% reporting moderate familiarity), while 3.4\% were neutral, and 6.1\% reported little to no prior experience. For frequency of AI chatbot use, 64.5\% used chatbots at least once a week (with 15.6\% using them daily, 28.6\% using them 3-5 times a week, and 20.2\% using them once a week), 29.8\% used them less frequently (13.7\% using them 2-3 times a month and 16.0\% using them once a month or less), and only 5.7\% did not use AI chatbots at all. All participation was voluntary, and the study received exempt approval from the University of Florida's Institutional Review Board.

\subsection{Analysis}

\begin{figure}
    \centering
    \includegraphics[width=\linewidth]{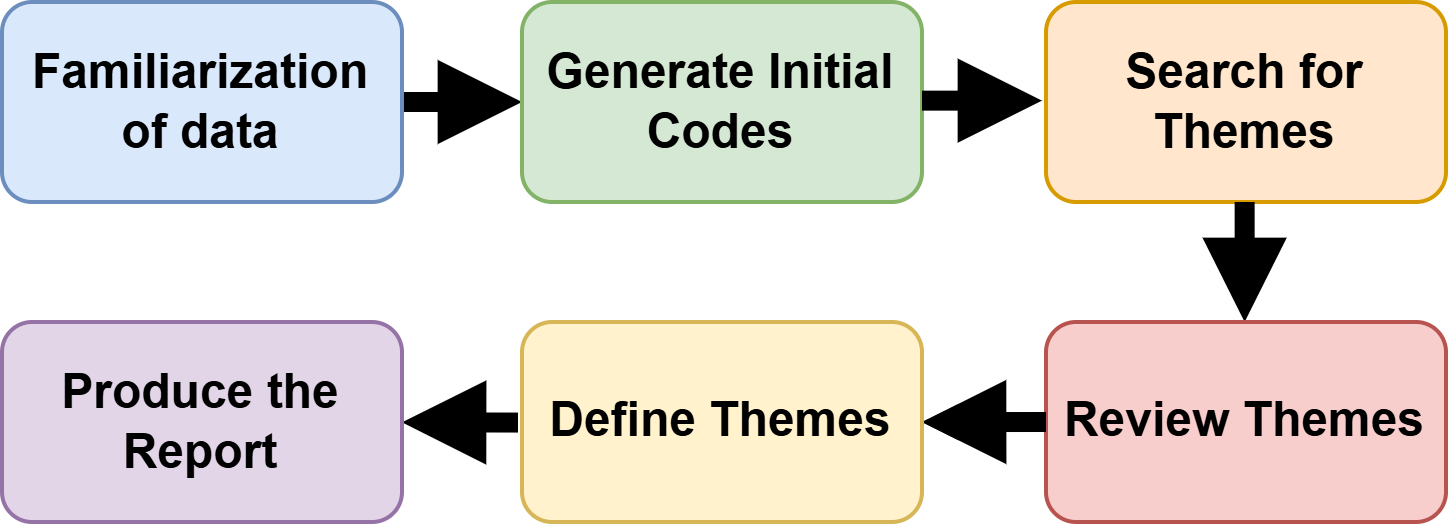}
    \caption{Six-phase Thematic Analysis}
    \label{fig:enter-label}
\end{figure}

The two open-ended questions about benefits and risks received a total of 745 distinct statements from the 262 participants, with most respondents providing multiple points in their answers. We analyzed these statements using Braun and Clarke’s \cite{30} six-phase thematic analysis framework: familiarization with data, generating initial codes, searching for themes, reviewing themes, defining and naming themes, and producing the report. Two researchers independently reviewed all responses multiple times to ensure thorough familiarization before inductively generating initial codes, allowing themes to emerge organically from the data rather than fitting responses into predetermined categories. Throughout the process, the researchers met regularly to discuss and refine the analysis, collaborating to define and consolidate themes. Any discrepancies in coding or theme identification were resolved through discussion until consensus was reached. Final themes were then carefully reviewed against the original data to ensure accurate representation of participant perspectives. Figure \ref{fig:enter-label} illustrates the complete thematic analysis procedure.

\section{Results}

Our thematic analysis of responses from 262 undergraduate participants yielded 745 distinct statements about AI chatbots in educational settings. The following sections present the themes from our analysis of perceived benefits and risks, which are summarized in Tables \ref{tab:benefits} and \ref{tab:risks} respectively.

\subsection{RQ1: Perceived Benefits of AI chatbots}

In response to the question ``If any, what do you see as the key benefits or opportunities for using AI chatbots in educational settings?'' five themes emerged from participant responses, with 5.5\% of students perceiving no benefits. These themes are summarized in Table \ref{tab:benefits} and explored in detail below.

\subsubsection{Feedback and Study Support}

Participants most frequently identified AI's ability to provide feedback and study support as a benefit. There were 92 statements relating to this theme, 23\% of the total. Students noted that AI chatbots are ``great for creating study guides, practice tests, organizing notes, and even breaking complicated concepts down to barebones for better understanding.” The technology's ability to ``help clear up misunderstandings or detect errors in content provided by students'' was valued, as was its capacity to generate ``infinite amounts of practice problems'' to assist in preparing for quizzes and exams. Students also appreciated AI's writing assistance features, which enabled them to double check grammar, proofread work, and otherwise assist in the ``editing process.''

\subsubsection{Instruction}

The ability of AI chatbots to function as personalized instructors emerged as the second most prominent theme, with 89 statements, 22.25\% of the total. Respondents noted that AIs can ``teach you things you are struggling with” by ``help[ing] with questions and [elaborating] on problems.” Respondents also stated that AI could provide ``personalized responses to specific questions” and ``customized explanation[s] of complex topics,” helping them understand information by summarizing key topics, explaining information in different ways, demonstrating ``step by step walkthroughs for challenging problems,” or otherwise providing additional context and definitions of terms. One respondent stated that they sometimes use AI to ``understand a difficult phrase by pasting it into the chatbot and asking it to explain it more clearly.” Respondents also noted the benefit of AI chatbots as a ``24/7 teacher” and ``free tutor” for students, allowing them ``quick tutoring opportunities” to ask questions if there was not enough time in class or office hours, or if ``a human teacher is not available.” AI also has the capacity for ``virtually unlimited follow up” questions.

\subsubsection{Access to Information}

Easy and efficient access to information constituted the third major benefit category, identified in 85 statements, or 21.25\% of the total. Within this theme, participants emphasized how AI chatbots provide rapid, reliable pathways to knowledge by ``[helping] understand key objectives [by] summarizing topics,” and directing users to relevant ``sources/articles for research.'' The technology's ability to offer ``clarification'' on challenging concepts by ``articulat[ing] it in a way that is easy to understand'' was frequently mentioned, along with practical applications like quickly retrieving ``formulas in Excel that [are not] committed to memory.” As one respondent stated, a primary advantage of AI ``include[s] getting an answer about a topic right away.'' Another respondent further described AI chatbots as ``ultimately [a] helpful tool for educational purposes. You have examples at your fingertips to do possibly anything.''

\begin{table*}[t]
\centering
\caption{Perceived Benefits of AI Chatbots in Educational Settings}
\label{tab:benefits}
\small
\setlength{\tabcolsep}{7pt}
\renewcommand{\arraystretch}{1.2}

\begin{tabularx}{\textwidth}{
@{}
>{\raggedright\arraybackslash}p{0.20\textwidth}
rr
>{\raggedright\arraybackslash}X
@{}
}
\toprule
\rowcolor{gray!15}
\textbf{Theme} & \textbf{\# of Statements} & \textbf{Percent} & \textbf{Example Quote} \\
\midrule

Feedback \& Study Support & 92 & 23.00\% &
``AI chatbots have the opportunity to create questions for you to test your knowledge in a variety of stem subjects.'' \\

\rowcolor{gray!5}
Instruction & 89 & 22.25\% &
``Chatbots can be great teachers on concepts that were not gone in depth about in class or the student needs more explanation to grasp the topic.'' \\

Access to Information & 85 & 21.25\% &
``Access and Availability are two key benefits. These systems can operate and be accessible around the clock and at any location that is accessible to the web (including a portable smart device)'' \\

\rowcolor{gray!5}
Productivity & 57 & 14.25\% &
``Save time and save a lot of unneeded tedious work.'' \\

Creative Support & 42 & 10.50\% &
``It provides a starting point for thought/creative processes, which can be used by students to further their research.'' \\

\rowcolor{gray!5}
No Benefit & 22 & 5.50\% &
``No benefits for AI in educational settings.'' \\

Other & 13 & 3.25\% & -- \\

\midrule
\rowcolor{gray!10}
\textbf{Total} & \textbf{400} & \textbf{100.00\%} & \\
\bottomrule

\end{tabularx}
\end{table*}

\subsubsection{Productivity}

Efficiency and productivity gains emerged as a benefit in 57 statements, accounting for 14.25\% of the total. Respondents noted that using AI can ``[save] time and energy” and ``accelerate menial, administrative tasks.” While researching, AI increases students’ efficiency by ``break[ing] down texts much quicker” and ``condens[ing] sources into short responses.” One respondent noted that ``it simplifies lessons, readings, and is easier and faster to use when it comes to research.” The technology's ability to provide targeted information was highlighted in observations that AI eliminates ``scrounging on Google to find a good explanation for a question...you get a direct response to your niche question.'' Beyond student applications, participants recognized potential faculty benefits, noting that AI can assist in developing ``assignment description[s] or...lesson plan[s]'' and ``help teachers generate homework and streamline the process of grading,'' allowing educators to ``spend more of their time focused on the students.''

\subsubsection{Creative Support}

AI's utility as an ideation tool and starting point generator was identified in 42 responses (10.50\% total). Participants valued how AI chatbots can help students ``get an idea as to how to start a certain problem...when they have no idea,'' facilitating the initial organization of thoughts and ``workshopping ideas and wording'' across academic tasks. This benefit extends, as one respondent noted that ``even if [the AI chatbot] is wrong, it can usually give you some sort of boost towards the correct answer or at least explain the problem well.'' In writing contexts specifically, participants appreciated AI's ability to ``[make] it easy to gather ideas and brainstorm for a paper'' and ``write up ideas or outlines,'' providing scaffolding for the creative process.

\subsubsection{No Benefit and Other}

Finally, 22 statements, accounting for 5.50\% of the total, explicitly stated they saw no educational benefits to AI chatbot implementation. These responses represent a counterbalance to the enthusiasm expressed in other themes, suggesting that despite the range of potential applications identified by most participants, a segment of the student population remains unconvinced of AI's educational value. The remaining 13 statements, or 3.25\% of the total, were coded as Other. These contained statements that were either too vague, incomplete, or general to categorize within the established themes. These included ambiguous statements like ``easier school life,'' incomplete thoughts, or minimal acknowledgments such as ``yes,'' ``some,'' or ``can help you.''

\subsection{RQ2: Perceived Risks of AI Chatbots}

In response to the question ``If any, what do you see as the main risks or concerns around using AI chatbots in educational settings?'' six distinct themes emerged from participant responses. These themes covered academic integrity concerns, information accuracy issues, potential skill degradation, potential overreliance, ethical considerations, and human connection losses. The following sections explore each theme in detail, as summarized in Table \ref{tab:risks}.

\subsubsection{Academic Integrity}

The largest category of statements included those concerned about academic integrity when using AI in educational settings. 97 statements were coded in this category, 28.12\% of the total. Specifically, respondents noted plagiarism and cheating as primary concerns. Typifying statements coded as academic integrity concerns, one respondent stated that, ``students might use AI chatbots to generate fully formed essays or assignments,” which ``is a serious risk.” The specific practice of students who ``copy and paste writing assignments from AI chatbots like Chat GPT and hand it in as their own original work'' was repeatedly highlighted as problematic. Many participants framed these behaviors within ethical and institutional contexts, noting that such actions ``are not only in violation of the school's honor code, but also unethical and unfair to their peers.'' Interestingly, a counterbalancing concern also emerged—that legitimate student work might be incorrectly flagged as AI-generated, with one participant noting they had ``already seen...peers...who do produce work that is entirely their own being falsely accused of submitting AI generated work.''

\begin{table*}[t]
\centering
\caption{Perceived Risks of AI Chatbots in Educational Settings}
\label{tab:risks}
\small
\setlength{\tabcolsep}{7pt}
\renewcommand{\arraystretch}{1.2}

\begin{tabularx}{\textwidth}{
@{}
>{\raggedright\arraybackslash}p{0.20\textwidth}
rr
>{\raggedright\arraybackslash}X
@{}
}
\toprule
\rowcolor{gray!15}
\textbf{Theme} & \textbf{\# of Statements} & \textbf{Percent} & \textbf{Example Quote} \\
\midrule

Academic Integrity & 97 & 28.12\% &
``Plagiarism -- some students copy and paste writing assignments from AI chatbots like Chat GPT and hand it in as their own original work.'' \\

\rowcolor{gray!5}
Accuracy of Information & 91 & 26.38\% &
``Could provide incorrect or misconstrued information/data on a given topic.'' \\

Loss of Critical Thinking & 67 & 19.42\% &
``It can lead to students not actually taking the time to learn certain subjects, instead using the AI chatbots to provide them with the answers they need.'' \\

\rowcolor{gray!5}
Overreliance & 29 & 8.41\% &
``Overreliance and trusting answers without question.'' \\

Ethical Concern & 23 & 6.67\% &
``AI chatbots aren't perfect and oftentimes can get even the most simple information incorrect. There are times where chatbots have been trained on improper data causing numerous such errors.'' \\

\rowcolor{gray!5}
Human Element & 18 & 5.22\% &
``They promote unoriginal ideas and a lack of creativity.'' \\

No Risk & 17 & 4.93\% &
``I don't see any risk.'' \\

\rowcolor{gray!5}
Other & 3 & 0.87\% & -- \\

\midrule
\rowcolor{gray!10}
\textbf{Total} & \textbf{345} & \textbf{100.00\%} & \\
\bottomrule

\end{tabularx}
\end{table*}

\subsubsection{Accuracy of Information}

Information reliability constituted the second most significant concern. 91 statements were coded in this category, 26.38\% of the total. Participants expressed wariness about misinformation, hallucinated data, and the absence of source citations in AI responses. One respondent stated that their main concern was ``the chatbot giving incorrect and fabricated information. If a student were to use one, and were given false information, they would not know it was false unless they put in the effort to cross reference it.” This reflects anxiety about AI-generated inaccuracies and students potential failure to verify such information, even if they are aware of the systems not being fully reliable. One respondent noted that AI chatbots ``do not always give accurate information as they are not ”aware” of what they are discussing. They are simply giving the best response to a series of words.”

\subsubsection{Loss of Critical Thinking}

Another major concern for respondents was students loss of critical thinking skills when using AI chatbots. This theme was found in 67 statements, 19.42\% of the total. Students noted concerns of fellow students no longer putting effort into properly learning information or completing assignments, simply having an AI tool generate a response for them. Respondents specifically noted that students might use conversation AI to ``avoid learning material” by asking chatbots for answers ``just . . . to complete an assignment, not to learn,” and ``it is very easy for students to easily find the answers without needing to actually understand the content.” One respondent even noted that ``students can pass hard courses [by] putting minimal effort and … using various AI tools. The student will not gain knowledge but will only cheat themselves.”

Students additionally identified specific cognitive and communication skills at risk, noting that AI dependence ``can lead to students not actually taking the time to learn certain subjects'' and might ``discourage independent thinking or research.'' Writing skill development received attention, with concerns that ``students might not develop good writing skills'' or could lose the ability ``to write independently.'' The worry centered on ``students using chatbots to avoid thinking, or putting effort into their assignments.''

This category of statements has a natural relation to academic integrity, with some respondents stating that, ``students may use chatbots as a way of cheating rather than actually learning the material,” and that ``students could just copy and paste responses straight from [the AI] so they did [not] learn anything about the particular prompt.

\subsubsection{Overreliance} 

Closely related to critical thinking concerns, yet distinct in its emphasis on dependency formation rather than skill erosion, overreliance on AI systems was noted in 29 statements (8.41\% of total). Several respondents stated their concern that students might use AI ``as a crutch, and lose the ability to fully think for” themselves, or simply that ``students may not think for themselves and rely on AI.” The potential for developing unhealthy solution patterns was noted, with one participant worried about students becoming ``[dependent] on AI chatbots to walk them through problems.'' Outside of cognitive impacts, participants identified potential character development issues, suggesting AI tools could ``encourage laziness,'' leading to students becoming ``over reliant'' and ``trusting [provided] answers without question.''

\subsubsection{Ethical Concerns}

Broader ethical considerations relating to the use of AI in education appeared in 23 statements (6.67\% of total), spanning algorithmic bias, labor market disruption, environmental impact, data privacy, and students’ mental health. Algorithmic bias received the most attention, with participants expressing anxiety about ``AI chatbots giving biased...information'' due to having ``been trained on improper data'' or providing ``limited information based on company beliefs.''

Data ethics formed another significant subset, with participants questioning information governance practices: ``it is not always known what the companies that run these will do with a user's data, where those companies got their training data and if they should have the rights to them, and how that data could possibly be misused.'' Environmental sustainability was referenced by multiple participants, with specific mention of energy consumption implications. Mental health impacts were also noted, with one participant expressing concern that ``users with mental health concerns may initiate off-topic emotional conversations that the chatbot is not prepared to answer, which can result in harm.''

\subsubsection{Human Element}

Concerns about diminished human presence and authentic expression in educational processes appeared in 18 statements, 5.22\% of the total. While some respondents were concerned simply about ``inauthentic work” or ``a lack of personality,” others noted that, ``especially in research papers or other pieces of literature, we lose a sense of humanity and reality in our writings when we use AI.” One respondent noted a concern that ``if using an AI chatbot for more than just outlining writing, it is very clear that the writing is AI and is taking away the opportunity for students to learn how to read critically, write eloquently, and grow a voice in the style of their writing.” Others were concerned about students losing the ability to be creative, original, and authentic if using AI chatbots. Standardization of expression was also flagged as problematic, with concerns that students might ``start having robot-like responses [and] nothing will be thought of organically'' or that AI ``could cause hundreds of kids to have the same written answer.''

Another concern respondents had was a lack of human interactions in learning, with participants troubled by AI ``removing the human element of transfer of knowledge'' and creating learning experiences that ``deviate from student-teach[er] interaction[s] and meaningful connections, [leading to a] less personalized education.'' Quality concerns were apparent in observations that AI provides ``lower quality of information and feedback as compared to human teachers.'' Some participants expressed anxiety about job displacement, noting that AI adoption ``could result in less availability of human teachers.'' The limitations of AI communication were also emphasized: ``[AI chatbots] are [not] as reliable as a human and can provide bad information. They also are [not] as flexible as a human (communication wise).''

\subsubsection{No Risk and Other}

Finally, 17 statements, 4.93\% of the total, responded as seeing no risk in students’ use of AI in educational settings. The remaining three statements (0.87\% of total) contained content that was either incomplete, unclear (such as the response ``unbalance''), or too general to categorize meaningfully, with responses limited to single affirmations like ``yes.''

\section{Discussion}

This study investigated student perspectives on AI chatbots in educational settings, addressing a gap in the literature where student voices have been underrepresented. Through a conducted survey and thematic analysis of 745 statements from 262 participants, we identified several themes relating to students' perceived benefits and risks of AI in education.

The most frequent benefit reported by students was feedback and study support. Students highlighted different ways AI supports their learning through its ability to provide immediate feedback. Identified use cases varied, from assistance with practice problems, to creating study guides, to proofreading and detecting errors in students' work. From a student perspective, this represents a direct and practical impact on their learning process, with students finding different ways to leverage available AI chatbots for study support, aligning with prior research on the benefits of immediate feedback and personalized learning support \cite{2,3,6,7,11,16,18,21}. The second most frequently stated benefit related to AI's potential capability to instruct. Unlike the feedback and study support theme where AI helps review or practice existing knowledge, students emphasized AI's ability to teach new content and explain unfamiliar concepts. Students valued that AI could ``teach you things you are struggling with,'' which from a student perspective, likened AI to a potential instructor or teaching assistant. This capability was valued during times when traditional academic support wasn't available, with AI serving as a ``24/7 teacher'' that could provide ``quick tutoring opportunities,'' aligning with prior research that highlighted the capability for AI chatbots to provide tutoring support to students \cite{1,2,6,7,21,16}. Information accessibility emerged as the third most prominent benefit, with students valuing AI's ability to provide ``an answer about a topic right away'' and access to information ``around the clock and at any location.'' This appreciation for continuous information access aligns with existing literature \cite{3,6,14,16,18} and represents another direct benefit for students. While prior research emphasized potential benefits for instructor and administrative tasks \cite{1,6,8,9,10,14,19,21}, students focused primarily on benefits directly impacting their own learning experiences, including support to productivity and creativity. Consideration should also be had for the approximate 5.5\% of students who wrote solely against the use of AI in educational settings.

The largest concern students identified with using AI chatbots in education was academic integrity. The emphasis on academic integrity aligns with prior studies \cite{2,17,18,21,23} that identified plagiarism concerns. However, students worry not only about intentional cheating but also about their legitimate work being mistakenly flagged as AI-generated, presenting a more complex challenge for instructors. The second most cited concern was accuracy of information. Unlike academic integrity, where the focus was on misuse, students emphasized concerns over the reliability of AI-generated information and assistance, aligning with concerns found in previous research \cite{10,19,51,29,25}. Students worried about ``the chatbot giving incorrect and fabricated information,'' and that without proper verification, students ``would not know it was false unless they put in the effort to cross reference it.'' The third most cited concern was loss of critical thinking skills. Students noted peers ``avoiding learning material'' by relying too heavily on AI, with which students might ``pass hard courses [by] putting minimal effort'' while not gaining actual knowledge. The concerns about skill degradation and overreliance align with and extend prior research \cite{2,10,18,23}, as students specifically responded with concerns to their diminishing ability to write and think independently.

The findings suggest many important considerations for implementing AI chatbots in educational settings. Educational institutions should establish clear policies addressing both ethical use and academic integrity concerns. These policies should include detailed guidelines for acceptable AI use across different academic contexts, from topic exploration to draft feedback, while addressing both plagiarism prevention and legitimate work authentication. Second, institutions should develop curriculum focused on teach students how to leverage AI tools effectively while maintaining their independence and critical thinking skills. This includes training on verifying AI-generated information, understanding AI's limitations, and using AI as a learning enhancement rather than a replacement for critical thinking. Third, institutions should proactively address emerging ethical challenges through data privacy measures, ensuring equitable access to AI tools, bias monitoring and mitigation strategies, and consideration of the environmental impacts associated with AI integration in educational contexts.

\subsection{Limitations and Future Work} 

While several insights are provided into student perspectives on AI chatbots in education, the limitations of this study suggest directions for future research. This study focused on undergraduate students at a limited number of institutions. Future studies could assess perspectives across different types of institutions, academic levels, and geographical regions to develop a more comprehensive understanding of student perspectives. Second, while this study found student concerns relating to potential loss of critical thinking skills and overreliance, understanding the long-term impacts of AI chatbots on learning outcomes requires additional research. Future studies could examine how patterns of AI usage affect student performance, skill development, and learning retention. Third, attitudes toward AI captured in this study represent a snapshot during a period of rapid technological advancement and shifting social acceptance of AI. Future longitudinal studies can examine how student perceptions change alongside technological advancements, changing institutional policies, and broader societal discourse around AI in education. Such research could help identify whether concerns about academic integrity and information accuracy persist as technologies mature and institutional policies adapt.

\section{Conclusion}

The increasing usage and development of AI chatbots in educational settings presents new benefits and risks for students and educators alike. While prior research has examined various aspects of opportunities and concerns relating to AI in education, there remained limited understanding of how students themselves view and experience these technologies. This study addressed this gap through a conducted survey and following thematic analysis to understand student perspectives on the benefits and risks of AI chatbots in higher education. 

The findings demonstrate alignment between student perspectives and prior research demonstrating potential the benefits of AI in education. Students valued the feedback and study support, instruction capabilities, increased access to information, increased productivity, increased creativity offered by AI chatbots. They specifically appreciated immediate assistance when instructors were unavailable, help with organizing tasks and brainstorming ideas, and clear explanations of complex topics. Students' concerns paralleled and expanded upon previous research. Academic integrity concerns extended beyond plagiarism to include worries about legitimate work being misidentified as AI-generated. Additional concerns related to the accuracy and reliability of information, the loss of critical thinking and writing skills, the potential development of overreliance, and ethical considerations regarding data privacy, system bias, environmental impact, and human interaction in education.

The successful implementation of AI chatbots in education requires careful consideration of both benefits and risks. By addressing student concerns, institutions and instructors can leverage AI towards areas such as immediate feedback and learning support. In the process, educational institutions should establish clear policies regarding AI use, develop curriculum that teaches skills necessary for effective AI use and critical thinking, and proactively address emerging ethical challenges including data privacy, system bias, and environmental impacts. This approach can enhance the quality of students' educational experiences while preserving the integrity of the learning process with AI.

\setcounter{enumiv}{0}
\bibliographystyle{unsrt}
\bibliography{acceptance}

@inproceedings{pitts3,
  title={Finding misleading identifiers in novice code using LLMs},
  author={{\v{R}}echt{\'a}{\v{c}}kov{\'a}, Anna and Maximova, Alexandra and Pitts, Griffin},
  booktitle={Proceedings of the 56th ACM Technical Symposium on Computer Science Education V. 2},
  pages={1595--1596},
  year={2025}
}

@inproceedings{a,
  title={Generation z and e-learning: Covid-19, personality, and technology acceptance},
  author={Marcus, Viktoria and Masrahi, Alaa and Motamedi, Sanaz},
  booktitle={Proceedings of the Human Factors and Ergonomics Society Annual Meeting},
  volume={66},
  number={1},
  pages={2103--2107},
  year={2022},
  organization={SAGE Publications Sage CA: Los Angeles, CA}
}

@inproceedings{b,
  title={Impact of Students’ Backgrounds on Online Learning Behavior: Generation Z Technology Acceptance of E-Learning Technology during COVID-19},
  author={Motamedi, Sanaz and Marcus, Viktoria},
  booktitle={2024 ASEE Annual Conference \& Exposition},
  year={2024}
}

@inproceedings{c,
  title={Understanding E-learning acceptance of gen z students: An extension of the technology acceptance model (TAM)},
  author={Motamedi, Sanaz and Marquis, Kierra and Levine, Hannah},
  booktitle={2021 ASEE Virtual Annual Conference Content Access},
  year={2021}
}

@inproceedings{1,
  title={Artificial intelligence technologies in education: benefits, challenges and strategies of implementation},
  author={Owoc, Mieczys{\l}aw L and Sawicka, Agnieszka and Weichbroth, Pawe{\l}},
  booktitle={IFIP International Workshop on Artificial Intelligence for Knowledge Management},
  pages={37--58},
  year={2019},
  organization={Springer}
}

@inproceedings{2,
  title={Programming is hard-or at least it used to be: Educational opportunities and challenges of AI code generation},
  author={Becker, Brett A and Denny, Paul and Finnie-Ansley, James and Luxton-Reilly, Andrew and Prather, James and Santos, Eddie Antonio},
  booktitle={Proceedings of the 54th ACM Technical Symposium on Computer Science Education V. 1},
  pages={500--506},
  year={2023}
}

@inproceedings{3,
  title={Investigating the benefits of applying artificial intelligence techniques to enhance learning experiences in capstone courses},
  author={A. Gonzalez, Luis},
  booktitle={Proceedings of the 17th ACM Conference on International Computing Education Research},
  pages={398--400},
  year={2021}
}

@inproceedings{4,
  title={An adaptive learning with gamification \& conversational UIs: The rise of CiboPoliBot},
  author={Fadhil, Ahmed and Villafiorita, Adolfo},
  booktitle={Adjunct publication of the 25th conference on user modeling, adaptation and personalization},
  pages={408--412},
  year={2017}
}

@article{6,
  title={Chatbots applications in education: A systematic review},
  author={Okonkwo, Chinedu Wilfred and Ade-Ibijola, Abejide},
  journal={Computers and Education: Artificial Intelligence},
  volume={2},
  pages={100033},
  year={2021},
  publisher={Elsevier}
}

@article{7,
  title={Towards applying powerful large AI models in classroom teaching: Opportunities, challenges and prospects},
  author={Tan, Kehui and Pang, Tianqi and Fan, Chenyou and Yu, Song},
  journal={arXiv preprint arXiv:2305.03433},
  year={2023}
}

@article{8,
  title={Teaching and Learning with AI-Generated Courseware: Lessons from the Classroom.},
  author={Schroeder, Kersten T and Hubertz, Martha and Van Campenhout, Rachel and Johnson, Benny G},
  journal={Online Learning},
  volume={26},
  number={3},
  pages={73--87},
  year={2022},
  publisher={ERIC}
}

@inproceedings{9,
  title={Student learning benefits of a mixed-reality teacher awareness tool in AI-enhanced classrooms},
  author={Holstein, Kenneth and McLaren, Bruce M and Aleven, Vincent},
  booktitle={Artificial Intelligence in Education: 19th International Conference, AIED 2018, London, UK, June 27--30, 2018, Proceedings, Part I 19},
  pages={154--168},
  year={2018},
  organization={Springer}
}

@article{10,
  title={Generative AI and teachers’ perspectives on its implementation in education},
  author={Kaplan-Rakowski, Regina and Grotewold, Kimberly and Hartwick, Peggy and Papin, Kevin},
  journal={Journal of Interactive Learning Research},
  volume={34},
  number={2},
  pages={313--338},
  year={2023},
  publisher={Association for the Advancement of Computing in Education (AACE)}
}

@inproceedings{11,
  title={Integrating ChatGPT in an Introductory Engineering Undergraduate Course as a Tool for Feedback},
  author={Cortez, Anthony and Schmelzenbach, Paul Daniel},
  booktitle={2024 ASEE Annual Conference \& Exposition},
  year={2024}
}

@inproceedings{12,
  title={Using AI Chatbots to Produce Engineering Spreadsheets in an Advanced Structural Steel Design Course},
  author={Campbell, Alexander},
  booktitle={2024 ASEE Annual Conference \& Exposition},
  year={2024}
}

@inproceedings{14,
  title={Creating and implementing a custom chatbot in engineering education},
  author={Abdulla, Shameel and Al Hamidi, Yasser M and Khraisheh, Marwan},
  booktitle={2023 ASEE Annual Conference \& Exposition},
  year={2023}
}

@article{15,
  title={Anne G. Neering: Interactive chatbot to engage and motivate engineering students},
  author={Crown, Stephen and Fuentes, Arturo and Jones, Robert and Nambiar, Rajiv and Crown, Deborah},
  journal={Computers in Education Journal},
  volume={21},
  number={2},
  pages={24--34},
  year={2011}
}

@inproceedings{16,
  title={How important are Chatbots within Engineering Education? A literature-based review from 2011 to 2023},
  author={Maladzhi, Rendani Wilson and Tsoeu, Mohohlo and Mthombeni, Nomcebo and Moloi, Katleho and others},
  booktitle={2023 World Engineering Education Forum-Global Engineering Deans Council (WEEF-GEDC)},
  pages={1--5},
  year={2023},
  organization={IEEE}
}

@article{17,
  title={Exploring the potential of generative artificial intelligence in education: applications, challenges, and future research directions},
  author={Hwang, Gwo-Jen and Chen, Nian-Shing},
  journal={Journal of Educational Technology \& Society},
  volume={26},
  number={2},
  year={2023},
  publisher={International Forum of Educational Technology \& Society}
}

@article{18,
  title={Reconceptualizing ChatGPT and generative AI as a student-driven innovation in higher education},
  author={Dai, Yun and Liu, Ang and Lim, Cher Ping},
  journal={Procedia CIRP},
  volume={119},
  pages={84--90},
  year={2023},
  publisher={Elsevier}
}

@article{19,
  title={Generative artificial intelligence in education, part one: The dynamic frontier},
  author={Hsu, Yu-Chang and Ching, Yu-Hui},
  journal={TechTrends},
  volume={67},
  number={4},
  pages={603--607},
  year={2023},
  publisher={Springer}
}

@article{21,
  title={The use of ChatGPT in the digital era: Perspectives on chatbot implementation},
  author={Limna, Pongsakorn and Kraiwanit, Tanpat and Jangjarat, Kris and Klayklung, Prapasiri and Chocksathaporn, Piyawatjana},
  journal={Journal of Applied Learning and Teaching},
  volume={6},
  number={1},
  pages={64--74},
  year={2023}
}

@article{23,
  title={The effects of over-reliance on AI dialogue systems on students' cognitive abilities: a systematic review},
  author={Zhai, Chunpeng and Wibowo, Santoso and Li, Lily D},
  journal={Smart Learning Environments},
  volume={11},
  number={1},
  pages={28},
  year={2024},
  publisher={Springer}
}

@inproceedings{25,
  title={A Proposed Model of Learners' Acceptance and Trust of Pedagogical Conversational AI},
  author={Pitts, Griffin and Marcus, Viktoria and Motamedi, Sanaz},
  booktitle={Proceedings of the Eleventh ACM Conference on Learning@ Scale},
  pages={427--432},
  year={2024}
}

@inproceedings{27,
  title={Gpteach: Interactive ta training with gpt-based students},
  author={Markel, Julia M and Opferman, Steven G and Landay, James A and Piech, Chris},
  booktitle={Proceedings of the tenth acm conference on learning@ scale},
  pages={226--236},
  year={2023}
}

@incollection{28,
  title={Jill Watson: A virtual teaching assistant for online education},
  author={Goel, Ashok K and Polepeddi, Lalith},
  booktitle={Learning engineering for online education},
  pages={120--143},
  year={2018},
  publisher={Routledge}
}

@article{29,
  title={Chatbots in education and research: A critical examination of ethical implications and solutions},
  author={Kooli, Chokri},
  journal={Sustainability},
  volume={15},
  number={7},
  pages={5614},
  year={2023},
  publisher={MDPI}
}

@article{30,
  title={Using thematic analysis in psychology},
  author={Braun, Virginia and Clarke, Victoria},
  journal={Qualitative research in psychology},
  volume={3},
  number={2},
  pages={77--101},
  year={2006},
  publisher={Taylor \& Francis}
}

@article{31,
  title={Perceptions and usage of AI chatbots among students in higher education across genders, academic levels and fields of study},
  author={St{\"o}hr, Christian and Ou, Amy Wanyu and Malmstr{\"o}m, Hans},
  journal={Computers and Education: Artificial Intelligence},
  volume={7},
  pages={100259},
  year={2024},
  publisher={Elsevier}
}

@inproceedings{40,
  title={Building trust between users and telecommunications data driven virtual assistants},
  author={Perez Garcia, Marta and Saffon Lopez, Sarita},
  booktitle={IFIP International Conference on Artificial Intelligence Applications and Innovations},
  pages={628--637},
  year={2018},
  organization={Springer}
}

@article{49,
  title={Transparency and the black box problem: Why we do not trust AI},
  author={Von Eschenbach, Warren J},
  journal={Philosophy \& Technology},
  volume={34},
  number={4},
  pages={1607--1622},
  year={2021},
  publisher={Springer}
}

@article{51,
  title={Unveiling security, privacy, and ethical concerns of ChatGPT},
  author={Wu, Xiaodong and Duan, Ran and Ni, Jianbing},
  journal={Journal of Information and Intelligence},
  volume={2},
  number={2},
  pages={102--115},
  year={2024},
  publisher={Elsevier}
}

\end{document}